%% file: main.tex
\documentclass[aps,amssymb,amsmath,pra,twocolumn,showpacs,superscriptaddress,longbibliography]{revtex4-2}

\usepackage{graphicx}
\usepackage{dcolumn}
\usepackage{bm}
\usepackage{subfigure}
\usepackage{color}

\usepackage{amssymb,amsmath,amsfonts,latexsym,graphicx,verbatim}

\usepackage[margin=0.5in]{geometry}

\usepackage[english]{babel}
\usepackage{times}
\usepackage{dsfont}
\usepackage{latexsym}
\usepackage{fancyhdr}
\usepackage{float}
\usepackage{afterpage}
\usepackage{enumitem}
\usepackage{mleftright}

\usepackage{eso-pic, graphicx}

\usepackage{listings}
\usepackage{multirow}
\usepackage{xcolor,colortbl}

\usepackage{bbm}
\usepackage{upgreek}
\usepackage{amsmath}
\usepackage{newtxtext,newtxmath}

\addto\captionsenglish{\renewcommand{\figurename}{Fig.}}

\definecolor{Ablue}{rgb}{0.96,0.24,0.00}

\definecolor{Abluetitle}{rgb}{0.,0.24,0.51}

\definecolor{orange}{rgb}{0.96,0.24,0.00}

\definecolor{darkred}{rgb}{0.55, 0.0, 0.0}

\newcommand{\red}[1]{\textcolor{red}{#1}}

\definecolor{darksalmon}{rgb}{0.91, 0.59, 0.48}
\definecolor{maroon}{cmyk}{0,0.87,0.68,0.32}

\definecolor{mustard}{rgb}{1.0, 0.86, 0.35}

\definecolor{Gray}{gray}{0.85}
\definecolor{LightCyan}{rgb}{0.88,1,1}
\newcolumntype{a}{$>${\columncolor{Gray}}c}
\newcolumntype{b}{$>${\columncolor{white}}c}

\usepackage{array}
\newcolumntype{L}[1]{$>${\raggedright\let\newline\\\arraybackslash\hspace{0pt}}m{#1}}
\newcolumntype{C}[1]{$>${\centering\let\newline\\\arraybackslash\hspace{0pt}}m{#1}}
\newcolumntype{R}[1]{$>${\raggedleft\let\newline\\\arraybackslash\hspace{0pt}}m{#1}}

\usepackage[colorlinks=true , citecolor=blue,urlcolor=blue]{hyperref}

\input{Commands3.tex}

\mleftright
\newcommand{\beginsupplement}{%
        \setcounter{table}{0}
        \renewcommand{\thetable}{S\arabic{table}}%
        \setcounter{figure}{0}
        \renewcommand{\thefigure}{S\arabic{figure}}%
				
     }

\newcommand{\affA}{Department of Chemistry, University of California, Berkeley, Berkeley, CA 94720, USA.}
\newcommand{\affB}{Chemical Sciences Division,  Lawrence Berkeley National Laboratory,  Berkeley, CA 94720, USA.}

\begin{document}
\title{Rapidly enhanced spin polarization injection in an optically pumped spin ratchet}

\author{Adrisha Sarkar}\thanks{Equal contribution}\affiliation{\affA}
\author{Brian Blankenship}\thanks{Equal contribution}\affiliation{\affA}
\author{Emanuel Druga}\affiliation{\affA}
\author{Arjun Pillai}\affiliation{\affA}
\author{Ruhee Nirodi}\affiliation{\affA}
\author{Siddharth Singh}\affiliation{\affA}
\author{Alexander Oddo}\affiliation{\affA}
\author{Paul Reshetikhin}\affiliation{\affA}
\author{Ashok Ajoy}\email{ashokaj@berkeley.edu}\affiliation{\affA}\affiliation{\affB}

\begin{abstract}
Rapid injection of spin polarization into an ensemble of nuclear spins is a problem of broad interest, spanning dynamic nuclear polarization (DNP) to quantum information science. We report on a strategy to boost the spin injection rate by exploiting electrons that can be rapidly polarized via high-power optical pumping. We demonstrate this in a model system of Nitrogen Vacancy center electrons injecting polarization into a bath of $\Cs$ nuclei in diamond. We deliver ${>}20$W of continuous, nearly isotropic, optical power to the sample, constituting a substantially higher power than in previous experiments. Through a spin-ratchet polarization transfer mechanism,  we show boosts in spin injection rates by over two orders of magnitude. Our experiments elucidate bottlenecks in the DNP process caused by rates of electron polarization, polarization transfer to proximal nuclei, and spin diffusion. This work demonstrates opportunities for rapid spin injection employing non-thermally generated electron polarization, and has relevance to a broad class of experimental systems including in DNP, quantum sensing, and spin-based MASERs.
\end{abstract}

\maketitle

\begin{figure}[t]
  \centering
  {\includegraphics[width=0.49\textwidth]{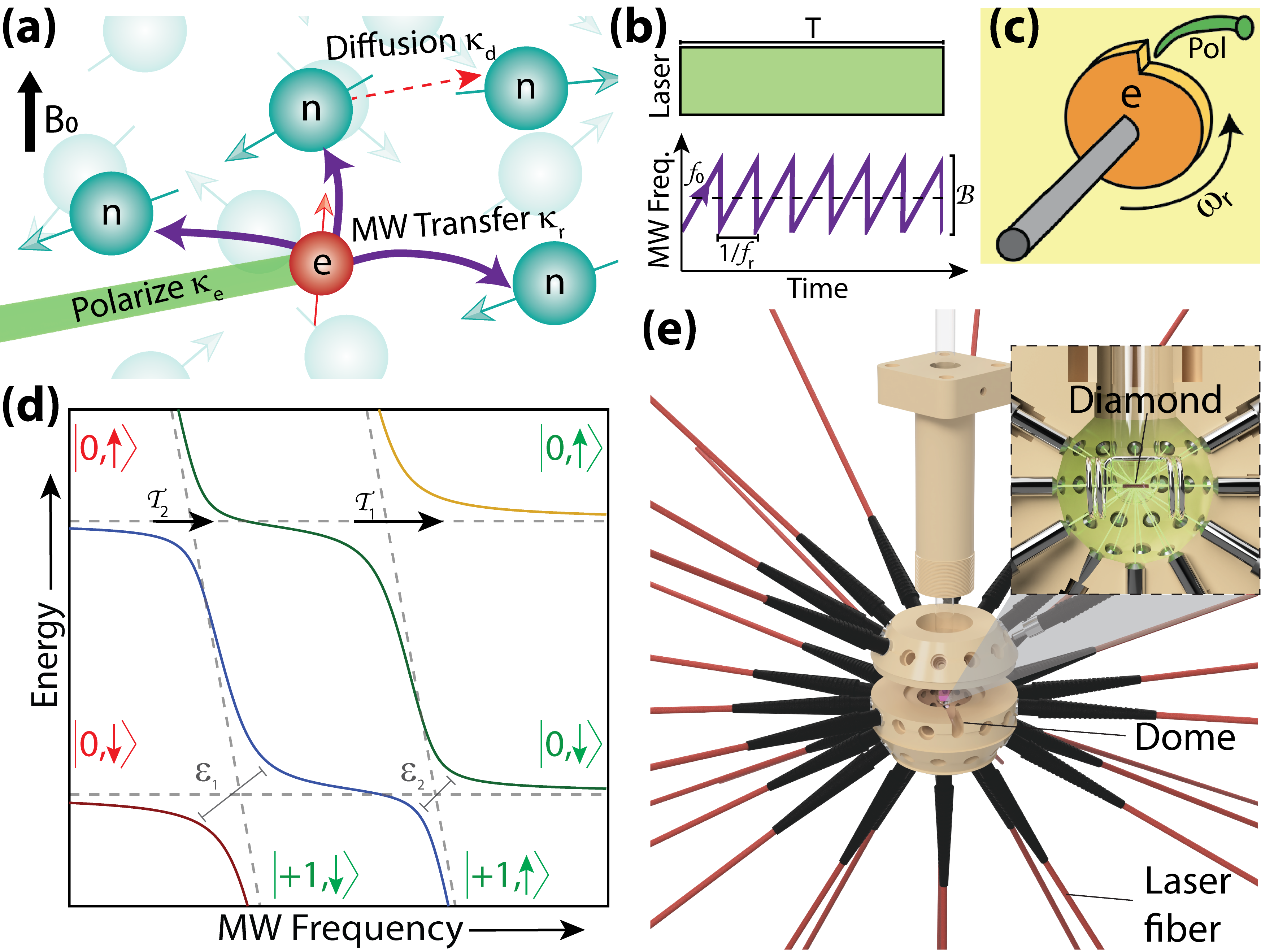}}
  \caption{\T{Spin ratchet $\Cs$ hyperpolarization.} (a) \I{Schematic lattice} consisting of an electron $e$ (red) and nuclei $n$ (blue). Hyperpolarization occurs through optical polarization of the electron (NV center), polarization transfer to directly-coupled nuclei ($\Cs$), and spin diffusion to bulk nuclei. Corresponding rate coefficients $\{\xk_e,\xk_r,\xk_d\}$ are depicted. (b) \I{DNP protocol} comprises continuous-wave (CW) laser and swept-MWs applied for period $T$. MW sweeps, at rate $f_r{=}\xo_r / 2\pi$, are applied over bandwidth $\mB$. (c) \I{Spin ratchet}. MW sweeps at rate $\xo_r$ drive optically generated $e$-polarization to nuclei akin to cranks of a ratchet. (d) \I{DNP mechanism}. Polarization transfer occurs via MW-driven traversals of cascaded Landau-Zener anti-crossings. Shown is the case for a single $e\tm n$ system (NV-$\Cs$). Energy gaps $\vxe_{1,2}$ are conditioned on the nuclear state (see \zar{ratchet}). (e) \I{Experimental setup}. Thirty ${\approx}0.8$W lasers (mounted on vertical walls) deliver optical excitation to the sample via optical fibers. Sample is positioned at the center, with lasers arranged in a dome shaped configuration.  \I{Inset:} Zoom into dome center showing diamond sample and split MW coil employed for hyperpolarization. }
\zfl{fig1}
\end{figure}

\section{Introduction}
\vspace{-4mm}
The injection of polarization into an ensemble of nuclear spins is a task of central importance in a variety of contexts. Not only is it the basis for dynamic nuclear polarization (DNP)~\cite{Abragam78,Goldman}, it is also important for the initialization of spin-based quantum information processors~\cite{Morton2008,Bluhm10,Bluhm11,Kloeffel13}, quantum sensors~\cite{Fuchs11,Reiserer16,Zaiser16}, and in emerging applications in spintronics~\cite{Reimer10,Yang18,Doherty16}. Critical to such applications is the \I{rate} at which polarization can be injected, measured, for instance, in terms of total angular momenta injected per unit time. For DNP applications, this rate ultimately determines the possible throughput of hyperpolarized spectroscopy and imaging~\cite{Hovener13}.

In this paper we demonstrate enhanced rates of spin injection employing \I{non-thermally} generated electron polarization~\cite{Reimer10,henstra88,Henstra90}. We focus on a model DNP system of {optically} polarizable Nitrogen Vacancy (NV) center electrons ($e$) in diamond and consider polarization transfer to lattice $\Cs$ nuclei ($n$)~\cite{London13,Fischer13}.  Considering \zfr{fig1}(a), the buildup of bulk nuclear polarization proceeds through a relayed process, and the overall rate of bulk spin injection is determined as an interplay between the rates at which $\I{(i)}$ electron polarization is generated, $\xk_e$, $\I{(ii)}$ polarization is transferred from electrons to proximal nuclei, $\xk_r$, and $\I{(iii)}$ polarization is relayed to distant (bulk) nuclei via spin diffusion, $\xk_d$ ~\cite{Redfield59}. In thermal DNP settings, $\xk_e{\app} T_{1e}^{-1}$ is system specific, cannot be controlled, and is relatively slow at low temperatures~\cite{Abragam82Book, Thurber10,Lund18}; likewise $\xk_r$ is constrained by the available microwave (MW) power at high magnetic fields~\cite{Becerra93,Maly08,Ramanathan08}. In contrast, in optical DNP, $\xk_e$ is determined by laser intensity,  potentially allowing access to the $\xk_e{\gg} T_{1e}^{-1}$ regime wherein source polarization can be rapidly generated.  Simultaneously, optical $e$-polarization can be carried out at low fields where high MW powers are readily accessible. In turn, this enables rapidly transmitting $e$-polarization to nuclei via high power MWs, significantly enhancing the rate of hyperpolarization.

A key contribution of this paper is the ability to deliver high optical power continuously to the electrons, allowing us to approach the $\xk_e{>}T_{1e}^{-1}$ regime (\zar{dome}). Time-averaged optical power here (${>}$20W) is substantially higher than previous experiments (\zar{table}).  We then demonstrate an approach to increase the polarization injection rate through a ``lockstep'' increase in laser and MW power at low fields. We identify hyperpolarization \I{rate-limits} by delineating experimental regimes where each of the three rates in \zfr{fig1}(a) individually bottleneck polarization injection. Ultimately, we show how high-power optical DNP can yield significant gains in polarization rate, boosted by as much as two orders of magnitude compared to conventional methods. Rapid hyperpolarization obtained here opens avenues for quantum sensors (e.g. gyroscopes~\cite{Ajoy12g,Ledbetter12} and spin sensors~\cite{Sahin21}) constructed out of hyperpolarized nuclear spins. Our approach  is also readily generalizable to other systems, including optically-pumped triplet molecular systems~\cite{Negoro18,Kouno20} and MASERs~\cite{Oxborrow12,Breeze18}.

\begin{figure*}[t]
  \centering
  {\includegraphics[width=0.95\textwidth]{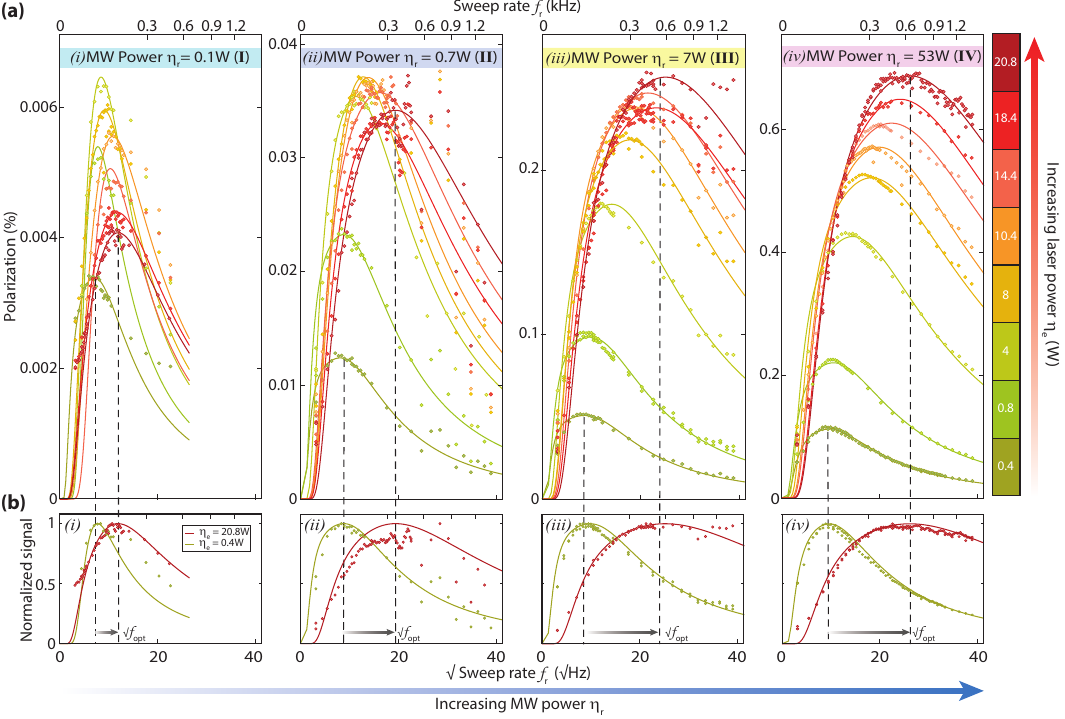}}
  \caption{\T{Enhancing rate of polarization transfer} at high laser and MW power. (a) DNP sweep rate profiles $P(f_r)$ for different regimes of MW power (\T{I-IV}) $\xn_r{=}$0.1W-53W, in each of which laser power is varied $\xn_e{=}$0.4W - 20.8W (see colorbar). Increase in $f_{\R{opt}}$ with increasing laser power is shown (dashed lines). Polarization time $T{=}$20s. Plots are shown against $f_r^{1/2}$ (upper axes denote $f_r$). Solid lines are fits to ratchet model (\zr{ratchet}). Corresponding values of maximum polarization and $f_{\R{opt}}$ are shown in \zfr{fig5}. Here, percent polarization was calculated through comparison with a thermal signal under the same experimental conditions and using the thermal Boltzmann distribution. (b) \I{Normalized curves} in (a) focusing on $\xn_e{=}$0.4W and 20.8W. Dashed lines elucidate optimal rates $f_{\R{opt}}$ (extended to (a)). Shifts are denoted by arrows. (\T{I-II}) At low MW powers, clustering of curves in (a) show that $f_{\R{opt}}$ changes slowly with $\xn_e$, yielding little polarization gain  (see \zfr{fig5}(b)-(c)).  (\T{III-IV}) At high MW powers, increasing $\xn_e$ shifts $f_{\R{opt}}$ to higher frequencies. Profiles in panels (b)\I{(iii-iv)} are almost identical, indicating that beyond a threshold, increasing MW power does not result in increasing $f_{\R{opt}}$ (see \zfr{fig5}(c)). }
\zfl{fig4}
\end{figure*}

 \I{System} --  In these experiments, we employ a single-crystal diamond sample with ${\sim}$1ppm NV concentration and natural abundance $\Cs$. The inter-NV spacing is ${\sim}$24nm~\cite{Reynhardt03a,Ajoy19relax}, and $\Cs$ lattice density is ${\sim}$0.92/nm$^3$. $T_{1n}{\app}$5min sets the overall polarization memory time for the system~\cite{Ajoy19relax}.  In what follows, we refer to the incident optical and MW power in Watts as $\xn_e$ and $\xn_r$ respectively. These are related to $\{\xk_e,\xk_r\}$ in \zfr{fig1}(a), but instead denote experimentally controllable parameters. 

DNP is excited at $B_0{=}36$mT through a mechanism involving CW optical and swept MW irradiation (\zfr{fig1}(b)) for period $T$~\cite{Ajoy17, Ajoy18}. NV centers are optically polarized to the $m_s{=}0$ state, yielding an $e$-polarization, $P_e(t){=}1-\exp(-t\xk_e)$, assuming a monoexponential rate constant $\xk_e$. In reality, $\xk_e{\app}c_e\xn_e$ is approximately proportional to the optical power applied. Precisely quantifying the absolute $e$-polarization is difficult due to dependence on several factors (e.g. interconversion between NV charge states~\cite{Grotz12,Aslam13}). However,  optically generated polarization obviates the need for high magnetic fields for hyperpolarization. Lower fields come with significant advantages, such as the availability of high MW powers at low frequencies. 

In our protocol, each MW sweep event injects polarization into $e$-proximal nuclei, akin to a \I{``ratchet''} (\zfr{fig1}(c)). Bulk polarization is then readout via RF induction at 7T. We employ pulsed spin-lock readout~\cite{Rhim76,Ajoy20DD} that permits interrogation of the $\Cs$ precession for multiple minute-long periods, with a decay constant $T_2^{\prime}{>}$20s~\cite{Beatrez21}. Signal is accumulated for the entire period and sampled every 1ns in windows between the pulses. As a result, we obtain high measurement fidelity, with integrated SNR as high as $10^8$ per shot~\cite{Beatrez21}.  We employ this high sensitivity to unravel the rate-limiting factors that affect polarization injection rates.

\I{Ratchet driven polarization} --  To elucidate the hyperpolarization mechanism, consider that the MW sweeps have an instantaneous frequency,  $f_{\R{MW}}(t){=}\mB f_r t+{f_0}-\mB/2$ (\zfr{fig1}(b)), where $\mB$ is the sweep bandwidth around $f_0$, the electronic spectral center, and $f_r$ is the sweep rate. $f_r$ plays a key role in setting the ultimate rate of bulk hyperpolarization buildup. Every sweep event transfers a finite amount of polarization, and the total number of sweeps, $T f_r$, is curtailed by the nuclear $T_{1n}$. Ideal ratchet operation involves maximizing $f_r$ while maintaining high polarization transfer efficiency \I{per} sweep.  Optimal operation occurs when the sweep rate $f_r{=}f_{\R{opt}}$, and can be derived by measuring how the polarization buildup rate $\dot P(f_r)$ depends on the interplay of rates in \zfr{fig1}(a) (see \zar{ratchet}).

Consider that the $e$-$n$ Hamiltonian for $N$ nuclei (\zfr{fig1}(a)) in the rotating frame at field $B_0$ has the form~\cite{Elanchezhian21},
\beq
\mH(\xo_{\R{MW}}){=}(\xD {-} \xg_eB_0)S_z^2 + \xO_{e}S_z + \xO_{e}S_x + \\
\sum_{j=1}^{N}\lsb \xo_j^{(0)}\mP_0I_{zj} + \xo_j^{(1)}\mP_1I_{z'j}\rsb
\zl{hamiltonian}
\eeq
where $\xo_{\R{MW}}{=}2\pi f_{\R{MW}}$, $S (I)$ refer to electron (nuclear) operators, and $\gamma_{e,n}$ are the respective magnetogyric ratios. The first two terms in \zr{hamiltonian} denote the NV zero-field splitting ($\xD{=}2.87$GHz) and Zeeman field; $\xO_e$ is the $e$-Rabi frequency ($\xO_e{=}c_r\xn_r$) proportional to the MW power applied, and we assume an NV axis aligned with $\T{B}_0$ (along $\hat{z}$). The last term in \zr{hamiltonian} describes nuclear field in either electronic manifold -- $\mP_0$ and $\mP_1$ are projection operators in the $m_s{=}0$ and $m_s{=}1$ manifolds, and the nuclear  resonance frequencies herein are $\xo^{(0)}_j$ and $\xo^{(1)}_j$ respectively~\cite{Elanchezhian21}: $\xo^{(0)}_j {\app} \xo_n {+} \fr{\xg_e B_0 A_{j}^{\pp}}{\xD}$, and, $\xo^{(1)}_j {=} [(\xo_n + A_{j}^{\pll})^2 + (A_{j}^{\pp})^2]^{1/2}$, where $\xo_n{=}\xg_nB_0$ and $\T{A}_j {=} A_{j}^{\pll}\hat{z} + A_{j}^{\pp}\hat{\pp}$ is the hyperfine coupling.

Spin ratchet polarization transfer is simplest to illustrate for a single $e\tm n$ system~\cite{Zangara18,Elanchezhian21}. Diagonalization leads to four Landau-Zener level anti-crossings (LZ-LACs), with energy gaps, $\vxe_{1}{\app}c_r\xn_r$ and  $\vxe_{2}{\app} \fr{2c_r\xn_r A_1^{\pp}}{\xo_n+A_1^{\pll}}$ (see \zfr{fig1}(d)), such that $\vxe_2{<} \vxe_1$ (here we set $\hbar{=}1$).  Swept MWs cause a traversal through this LZ-LAC cascade. Its action can be evaluated under simplifying approximations that capture the experiments~\cite{Elanchezhian21}: \I{(i)} LZ-LACs are assumed to be traversed sequentially, and, \I{(ii)} $e$-repolarization is assumed to occur at the start of every sweep event. This entails negligible laser action at the LZ-LACs, valid when $\mB{\gg} \vxe_{1,2}$, as in our experiments. Hyperpolarization is generated because the energy gaps are conditioned on the nuclear state; traversals through the LZ-LACs are differentially adiabatic or diabatic, leading to a population bias towards one nuclear state (here $\ket{\dw}$). Population bifurcation at an LZ-LAC is set by its adiabaticity and captured by respective {tunneling} probabilities $\mT_{1,2}(f_r) {=}\exp(-\vxe_{1,2}^2/f_r\mB)$ (\zfr{fig1}(d)). Hyperpolarization occurs when sweep rates $f_r$ are such that $\mT_1{\rt}0$ (adiabatic) and $ \mT_2{\neq} 0$ (diabatic)~\cite{Elanchezhian21}. The rate of polarization buildup is then (see \zar{ratchet}), 
\beq
\dot P(f_r) \app f_r\left[1-\exp \left(\frac{-c_e\xn_e}{f_{r}}\right)\right]\left[\left(1-\mT_{2}\right)\lcb 1-\left(2 \mT_{1}-1\right)^{2}\rcb \right] \:.
\zl{ratchet}
\eeq
The first term in square brackets is the rate of $e$-polarization and the second term captures the $e{\rt}n$ polarization transfer efficiency per sweep. An optimal rate, $f_{\R{opt}}$,  occurs when $d\dot P/df_r{=}0$. 

It is intuitive to see why an optimal rate $f_{\R{opt}}$ should exist. More rapid sweeps (high-$f_r$) can allow faster ratchet operation, but comes at the cost of \I{(i)} reduced electron polarization, and, \I{(ii)} lower transfer efficiency per sweep because the differential adiabaticity in \zfr{fig1}(d) is compromised. At very high-$f_r$, $\mT_1 {\app} \mT_2 {\app} 1$, and $\dot P{\rt}0$. Therefore $f_{\R{opt}}$ reflects an overall rate-limit for $e{\rt}n$ polarization transfer, and defines the rate at which the ratchet (\zfr{fig1}(c)) should be operated. The physics of direct $e{\rt} n$ polarization transfer is essentially unchanged from \zr{ratchet} even for large $N$~\cite{Elanchezhian21, Pillai21}; however, since experiments primarily probe \I{bulk} nuclei, there is a non-negligible role from spin diffusion.

Importantly, \zr{ratchet} suggests that the rate of polarization transfer, encoded in $f_{\R{opt}}$, can be enhanced through a simultaneous increase in laser and MW power -- higher $\xn_e$ allows faster $e$-repolarization, and higher $\xn_r$ permits faster sweeps while maintaining differential adiabaticity. This is the key insight we exploit in this work. In the ultimate limit, the spin ratchet hyperpolarization rate (encoded by $f_{\R{opt}}$) is only limited by spin diffusion.

\I{Enhanced optical excitation} -- Polarization rates here can exceed that of conventional DNP because electrons can be optically polarized faster than $T_{1e}^{-1}$, and this polarization can be rapidly transferred to proximal nuclei. We develop a laser delivery apparatus that enables continuous sample irradiation at $\xn_e{>}20$W. The apparatus (\zfr{fig1}(e)) design and construction is detailed in \zar{dome}. It consists of a 3D printed dome-shaped structure  (``laser dome'') that houses 30 multimode fibers attached to diode lasers ($\approx$0.8W each). To manage sample heating associated with high optical powers, we designed an \I{in-situ} heat exchanger.  Its performance (detailed in \zar{dome}) is highly efficient: even at sustained 24W optical power (an intensity of ${\app}0.19$kW/cm$^2$) employing thirty lasers, the steady-state temperature reached in the sample vicinity is less than $30^{\degree}$C.

\begin{figure}[t]
  \centering
  {\includegraphics[width=0.49\textwidth]{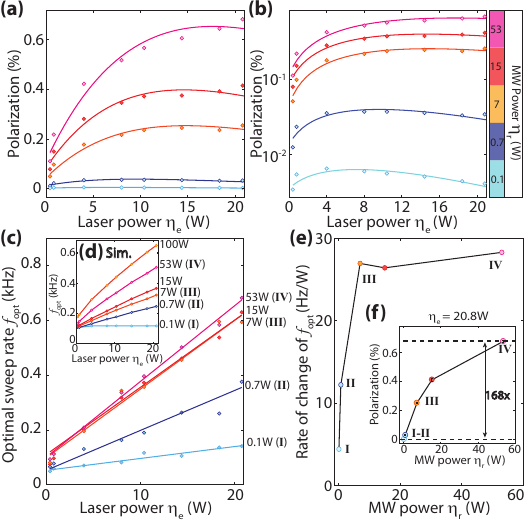}}
    \caption{\T{Speed limits for polarization transfer. } (a-b) \I{Maximum polarization level achieved} in \zfr{fig4} plotted on a (a) linear and (b) logarithmic scale. Curves correspond to the $\xn_r$-regimes studied. Combined optical-MW $\xn_e\tm\xn_r$ power increase yields a ${>}$200-fold boost in hyperpolarization level. (c) \I{Optimal sweep rates} $f_{\R{opt}}$ extracted from \zfr{fig4} for increasing $\xn_e$, at different MW power regimes \T{(I-IV)}. Points are extracted $f_{\R{opt}}$ values, while lines are linear fits. (d) \I{Spin ratchet simulations} similar to (c). Slopes here continually increase with $\xn_r$ due to absence of spin diffusion. (e) \I{Speed up of ratchet} $df_{\R{opt}}/d\xn_e$ obtained from slopes in (c). Plateau at high MW power indicates a spin diffusion bottleneck. (f) \I{Polarization increase with MW power}. Vertical slice of data in (a) at  $\xn_e{=}$20.8W. Polarization begins to plateau with increasing MW power.}
\zfl{fig5}
\end{figure}

\begin{figure}[t]
  \centering
  {\includegraphics[width=0.49\textwidth]{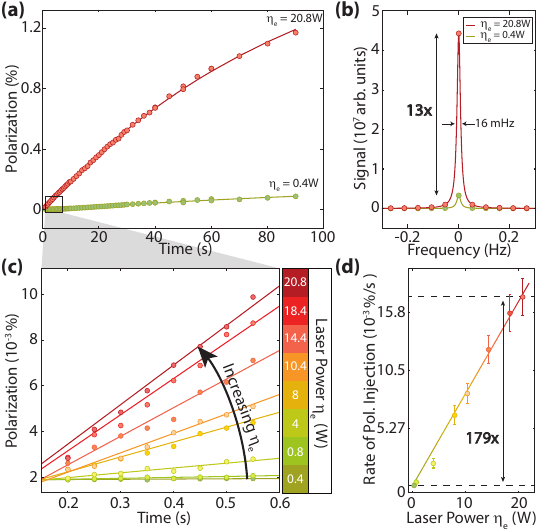}}
  \caption{\T{Enhanced spin injection via high-power optical pumping.} (a) \I{DNP buildup curve} in the high-MW power regime \T{IV}, $\eta_r{=}$53W with laser power $\xn_e{=}$0.4W and 20.8W. There is rapid polarization growth in the latter case. (b) \I{Corresponding NMR spectra} at $t{=}$90s obtained via pulsed-spin locking. Signal is enhanced ${\approx}$13-fold for $\xn_e{=}$20.8W compared to 0.4W optical power. DNP signal at $\xn_e{=}$ 20.8W corresponds to 1.17\% polarization, a ${>}10^{3}$ enhancement over thermal signal. (c) \I{Short-time $\Cs$ polarization growth}  as a function of laser power $\xn_e$, obtained by linearizing the $T{<}$0.6s portion of the buildup curves. Colors represent different laser powers $\eta_e$ (colorbar). Solid lines are linear fits. (d) \I{Extracted polarization injection rates} from (c) show an approximately linear increase with laser power. We obtain a ${\app}$179-fold boost in polarization injection when applying 20.8W of optical power}
\zfl{fig6}
\end{figure}

\section{Results}
\vspace{-4mm}
\I{Lockstep polarization increase} -- In the ratchet model without spin diffusion, $f_{\R{opt}}$ plays a key role as a proxy for the rate of nuclear spin polarization injection.  We now demonstrate that high power optical delivery via the apparatus in \zfr{fig1}(e), combined with simultaneous high MW power, can yield significant enhancements to $f_{\R{opt}}$, and thereby, spin injection rates. In parallel, we quantify the rate-limits for bulk polarization buildup as an interplay of the rates discussed in \zfr{fig1}(a).

Consider first data in \zfr{fig4}. We measure sweep rate dependent DNP profiles $P(f_r)$ with $T{=}20$s, at several regimes of optical $\xn_e$ and MW $\xn_r$ power. In particular, we consider four regimes of MW power between $\xn_r{=}0.1\tm 53$W, and for each, measure $P(f_r)$ profiles for different effective optical powers $\xn_e{=}$0.4W-20.8W. Upper and lower axes in \zfr{fig4} denote $f_r$ and $f_r^{1/2}$ respectively. Solid lines are fits to \zr{ratchet}. Optimal sweep rates $f_{\R{opt}}$ (dashed lines) are determined from maxima of the fitted curves.

\zfr{fig5}(a)-(b) show obtained maximal polarizations against $\xn_e$ for the MW power regimes considered (colorbar), plotted on a linear (\zfr{fig5}(a)) and  logarithmic (\zfr{fig5}(b)) scale. Polarization here is generated at $f_{\R{opt}}$. Solid lines are guides to the eye. In a complementary manner, points in \zfr{fig5}(c) display the extracted optimal rates $f_{\R{opt}}(\xn_e)$ for different MW powers. Solid lines are linear fits. \zfr{fig5}(e), in turn, plots the slope of these lines, ${\sim}df_{\R{opt}}(\xn_e)/d\xn_e$, while \zfr{fig5}(f) plots the polarization generated using different MW powers at $\xn_e{=}20.8$W (vertical slice in \zfr{fig5}(a)). A combined view of \zfr{fig4}-\zfr{fig5} allows the ability to correlate $f_{\R{opt}}$ to the absolute spin injection rates achieved.

We now systematically consider factors setting $f_{\R{opt}}$ from left-to-right in \zfr{fig4}, starting from the low-MW power region $\T{I}$ ($\xn_r{=}0.1$W). This regime relates to the situation in high-field DNP where MW power is low due to technological constraints. In \zfr{fig4}\I{(i)}, we observe the DNP profiles clustered at $f_r{=}$50-120Hz. Increasing optical power only weakly increases $f_{\R{opt}}$ (see also \zfr{fig5}(c)\T{(I)}), and there is little increase in hyperpolarization.  This indicates a $\xk_r$ induced speed-limit set by adiabaticity constraints -- in the ratchet picture in \zfr{fig1}(d), the energy gaps are small and sweep rates $f_r$ required to satisfy adiabaticity are slow.  Increasing optical power therefore yields no significant increase in nuclear polarization levels (see \zfr{fig4}(a)\I{(i)}). We observe in fact, a slight \I{decrease} in polarization at high $\xn_e$ (see also \zfr{fig5}(b)), associated with the rightward shift in $f_{\R{opt}}$. This polarization drop is beyond the scope of \zr{ratchet} and challenging to model. We hypothesize it arises because the electrons are repolarized multiple times during each MW sweep event, including at the LZ-LACs, making polarization transfer less efficient. Simultaneously, we observe oscillations in the $P(f_r)$ profiles in the high $\xn_e\tm\xn_r$ regime.

Now upon increasing MW power seven-fold (\zfr{fig4}\I{(ii)}) to regime \T{II} with $\xn_r{=}$0.7W, there is a larger increase in $f_{\R{opt}}(\xn_e)$ with optical power. This rightward shift in \zfr{fig4}(b)\I{(ii)} manifests as an increased slope in \zfr{fig5}(c)\T{(II)}. Intuitively, higher $\xn_r$ yields larger energy gaps in \zfr{fig1}(d) and allows faster sweep rates. When the optical power is low, $e$-polarization is not produced rapidly enough at the NV source. Increasing $\xn_e$ relieves this $\xk_e$ bottleneck. From \zfr{fig5}(e), $f_{\R{opt}}$ in regime \T{II} increases ${\sim}$2-fold with respect to regime \T{I}, and there is a simultaneous ${\sim}10$x increase in polarization (\zfr{fig5}(b)). However, clustering of $P(f_r)$ profiles in \zfr{fig4}\I{(ii)}, and slow growth in \zfr{fig5}(c)\T{(II)}, suggests that $\xk_r$ still limits the rate of polarization transfer.

 \zfr{fig4}\I{(iii)} considers a further 10-fold increase in MW power (regime \T{III}). Here a rightward shift in $f_{\R{opt}}$ is clearly evident with increasing optical power, and spin-ratchet operation is rapid, evidenced by the increased slope in \zfr{fig5}(c)\T{(III)}. Higher $\xn_r$ allows faster sweeps due to weaker adiabaticity constraints, and $\xn_e$ can be simultaneously boosted to increase the rate of source $e$-polarization, yielding a \I{lockstep} $\xk_e\tm\xk_r$ increase in hyperpolarization rate. The concomitant polarization increase is evident in \zfr{fig5}(a)-(b). We also note that the resulting maximal rates $f_{\R{opt}}{\app}0.65$kHz approach, and potentially exceed, the thermal rate $T_{1e}^{-1}{\sim} 0.2$kHz~\cite{Jarmola21} expected in this sample.

One might expect that a further increase in MW power will continue to yield such gains. This is shown in simulations in \zfr{fig5}(d) where we model direct $e{\rt}n$ polarization transfer without spin diffusion. Interestingly, however, we experimentally observe that a subsequent increase in MW power to $\xn_r{=}$53W (regime \T{IV}) provides no significant increase in $f_{\R{opt}}$ (see \zfr{fig4}\I{(iv)}). This is also reflected in \zfr{fig5}(c), where $f_{\R{opt}}(\xn_e)$ manifests as a series of overlapping lines beyond $\xn_r{=}$7W. This plateauing of rates is also evident in \zfr{fig5}(e). Correspondingly, the relative polarization increase in \zfr{fig5}(a)-(b) begins to slow down (see \zfr{fig5}(f)). Overall this points to the presence of a third speed-limit, which we ascribe to spin diffusion $\xk_d$. Here polarization is rapidly transferred from the NV center to proximal $\Cs$ nuclei, but is limited in its ability to reach the bulk nuclei. Indeed,  \zfr{fig5}(b)-(c) demonstrates that while $e$-polarization rates increase with $\xn_e$,  nuclear spin injection is ultimately limited by spin diffusion. As such, spin injection can be considered optimally rapid in this regime.

Finally, let us quantify spin injection gains from our strategy with respect to more conventional approaches. We attempt to make two comparisons: first, gains with respect to typical optical DNP experiments (see~\zar{table}), and second, gains with respect to the regime typically employed in high-field DNP. For the first case, consider spin injection with two optical powers, $\xn_e{=}$0.4W and 20.8W in regime \T{IV}. The former is representative of powers employed previously for optical DNP~ \cite{Eichhorn13,Tateishi14} (\zar{table}), and the latter corresponds to the use of the apparatus in \zfr{fig1}(e). \zfr{fig6}(a) shows the respective polarization buildup curves. The steeper polarization growth in the latter is evident. \zfr{fig6}(b) shows the corresponding $\Cs$ NMR spectra for $T{=}90$s, wherein an ${\app}$13-fold increase in signal is observed. Narrow spectral linewidths ${\app}$16mHz here are due to slow spin-lock decays~\cite{Beatrez21}. To focus on $e$-proximal polarization injection separately from the strong effects of spin diffusion, \zfr{fig6}(c) instead considers the small-time regime ($T{<}0.6$s). Polarization buildup is approximately linear in this regime; colorbar shows different optical powers employed. Corresponding slopes (\zfr{fig6}(d)) permit quantification of absolute polarization injection rates (\%/s) by comparing against the thermal polarization at 7T (${\sim}10^{-5}$). \zfr{fig6}(d) demonstrates that spin injection rate scales approximately linearly with optical power, arising from the increased rate of source $e$-polarization $\xk_e$ at the NV center sites. Indeed, sustained high-power optical delivery yields a ${\app}$179-fold increase in polarization injection with higher power pumping. In this case, we measure a linearized bulk injection rate of ${\app}$0.016\%/s averaged over the sample.

For the second comparison with respect to conventional DNP, consider two MW powers, $\xn_r{=}$0.1W and $\xn_r{=}$53W (regimes \T{I} and \T{IV} respectively) at $\xn_e{=}$20.8W. The former is representative of typical MW powers in high-field DNP (without gyrotrons). From the peaks of the corresponding traces in \zfr{fig4}, we observe an ${\app}$179-fold increase in polarization in the latter case (see \zfr{fig5}(f)). Overall, while a precise comparison between disparate experimental systems is difficult, data in \zfr{fig4} and \zfr{fig6} indicate that harnessing non-thermal $e$ polarization can yield spin injection gains by more than two orders of magnitude over conventional approaches.
\begin{figure*}[t]
  \centering
  {\includegraphics[width=0.98\textwidth]{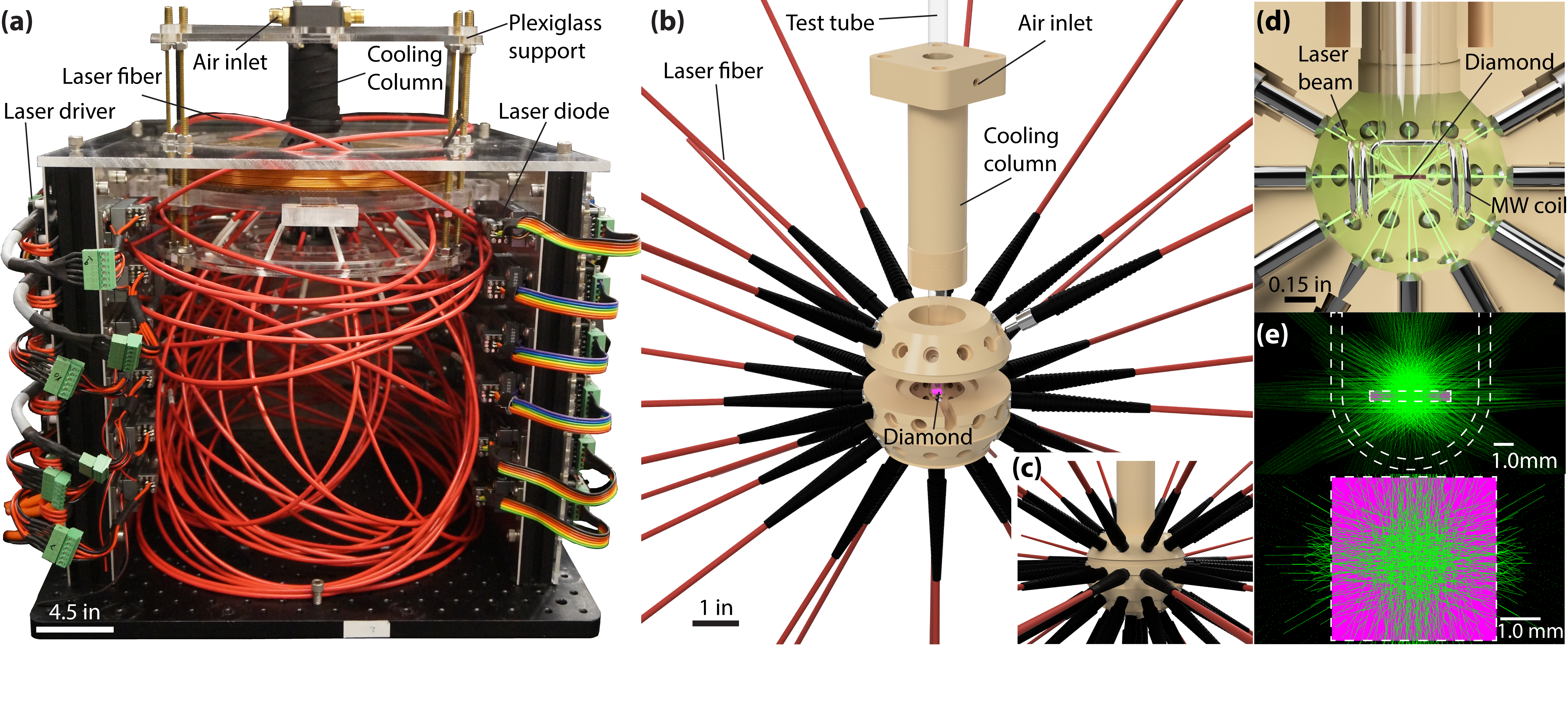}}
  \caption{\T{Setup for high power optical hyperpolarization. } (a) \I{Photograph.} Thirty ${\approx}0.8$W lasers (mounted on vertical walls) deliver optical excitation to the sample via optical fibers. Sample is positioned at the center, with lasers arranged in a dome shaped configuration. (b) \I{Laser dome} comprises 3D printed holder for optical illumination (shown is an expanded view). Centrally aligned bores attach optical fibers via a pressure fit. Diamond is held in a test tube at the center. Neck shaped region at the top houses a cooling column for heat exchange (see \zfr{fig3}(a)). (c) Rendering of dome with lasers attached. Corresponding implementation is shown in (a). (d) \I{Isotropic excitation}. Zoomed image of the dome center depicting isotropic laser illumination onto diamond sample. Split MW coil employed for hyperpolarization is shown. (e)\I{ Geometric optics simulations} showing sample illumination from side and top views. Only refraction is modeled (attenuation is neglected). }
	\zfl{fig2}
\end{figure*}

\begin{figure*}[t]
  \centering
  {\includegraphics[width=0.96\textwidth]{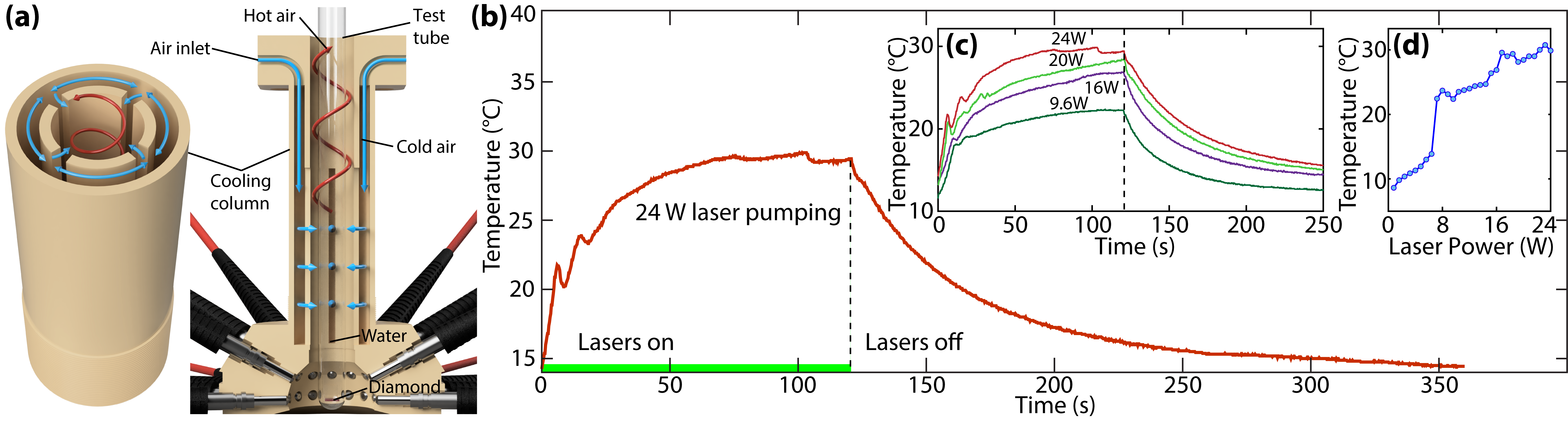}}
  \caption{\T{Thermal management for high power optical pumping. } (a) \I{Sample cooling strategy}. Cross-section of the laser dome neck (see \zfr{fig2}(b)) (top and side views). -20$^{\degree}$C N$_2$ gas (blue arrows) flows through the air inlets, and actively cools the walls of the test tube carrying water that surrounds the sample. Blue and red arrows depict cold and hot air currents respectively. (b) \I{Heat exchanger performance.} Profile of heat buildup and dissipation upon 120s of continuous 24W optical illumination from 30 lasers and subsequent turn off (dashed vertical line). Temperature is measured 4mm away from the diamond sample in the test tube. Steady-state temperature does not exceed 30$^{\degree}$C even with a sustained 2min high-power illumination. (c) \I{Temperature buildup and cooling profiles} for different optical powers. Sample cooling upon laser turnoff (dashed line) occurs with an exponential time constant of $\approx$1min. (d) \I{Measured steady-state temperature} (after 120s) for varying optical power applied. }
\zfl{fig3}
\end{figure*}

\vspace{-8mm}
\section{Discussion}
\vspace{-4mm}
\I{Applications} -- The experiments here illustrate the strength of utilizing non-thermally polarized electrons for DNP. $e$-polarization can be replenished at a rate $\xk_e{>} T_{1e}^{-1}$, and does not require cryogenic high-field conditions. This allows access to low-field regimes where MW power can be plentiful.  Simultaneously applied high optical and MW power can yield bulk spin injection rates ultimately limited by only spin diffusion. Similar arguments can be extended to other non-thermally generated DNP approaches, including with parahydrogen~\cite{Hovener13,Adams09,Pravdivtsev213}.

The rapidly injected spin polarization here projects onto applications that exploit hyperpolarized $\Cs$ spins as quantum sensors~\cite{Sahin21}, leveraging their long lifetimes in the laboratory~\cite{Ajoy19relax,Rej15} and rotating frames~\cite{Ajoy20DD,Beatrez21}.  A particularly compelling application is in nuclear spin gyroscopes; while previous work proposed~\cite{Ledbetter12,Ajoy12g,Maclaurin12} and demonstrated~\cite{Jaskula19,Soshenko21,Jarmola21} gryoscopes constructed from polarized ${}^{14}$N nuclear spins in diamond, the use of rapidly polarized $\Cs$ nuclei and their relatively high spin density promises to yield significant sensitivity boosts. A similar application is the use of hyperpolarized $\Cs$ nuclei as magnetometers, especially in high-field settings~\cite{Sahin21}; rapid hyperpolarization plays a critical role in enabling high sensitivity. We envision other applications in sensors for dark-matter searches~\cite{Aybas21,Garcon17},  RF imaging agents~\cite{Lv19},  nuclear spin RASERs~\cite{Oxborrow12,Breeze18,Appelt19}, and in condensed matter applications probing driven non-equilibrium states of matter~\cite{Beatrez22}.

\I{Conclusions} -- In conclusion, we have demonstrated the ability to rapidly inject spin polarization into a lattice of nuclear spins, via optically polarized electrons under simultaneous high-power optical and MW irradiation. In the process, we uncovered rate-limits that bottleneck bulk polarization transfer in various regimes, and showed that high power excitation can yield gains in spin-injection rate that exceed two orders of magnitude. Our work informs on interesting opportunities afforded by non-thermally polarized electrons for DNP and quantum sensing.

\section{Acknowledgements}
\vspace{-4mm}
We gratefully acknowledge M.  Markham (Element6) for the diamond sample used in this work, and discussions with J. Reimer, C. Meriles and D.Suter. This work was funded by ONR under N00014-20-1-2806. B.B. acknowledges support from a NSF Graduate Research Fellowship (DGE 2146752)

\begin{figure*}
  \centering
  {\includegraphics[width=0.9\textwidth]{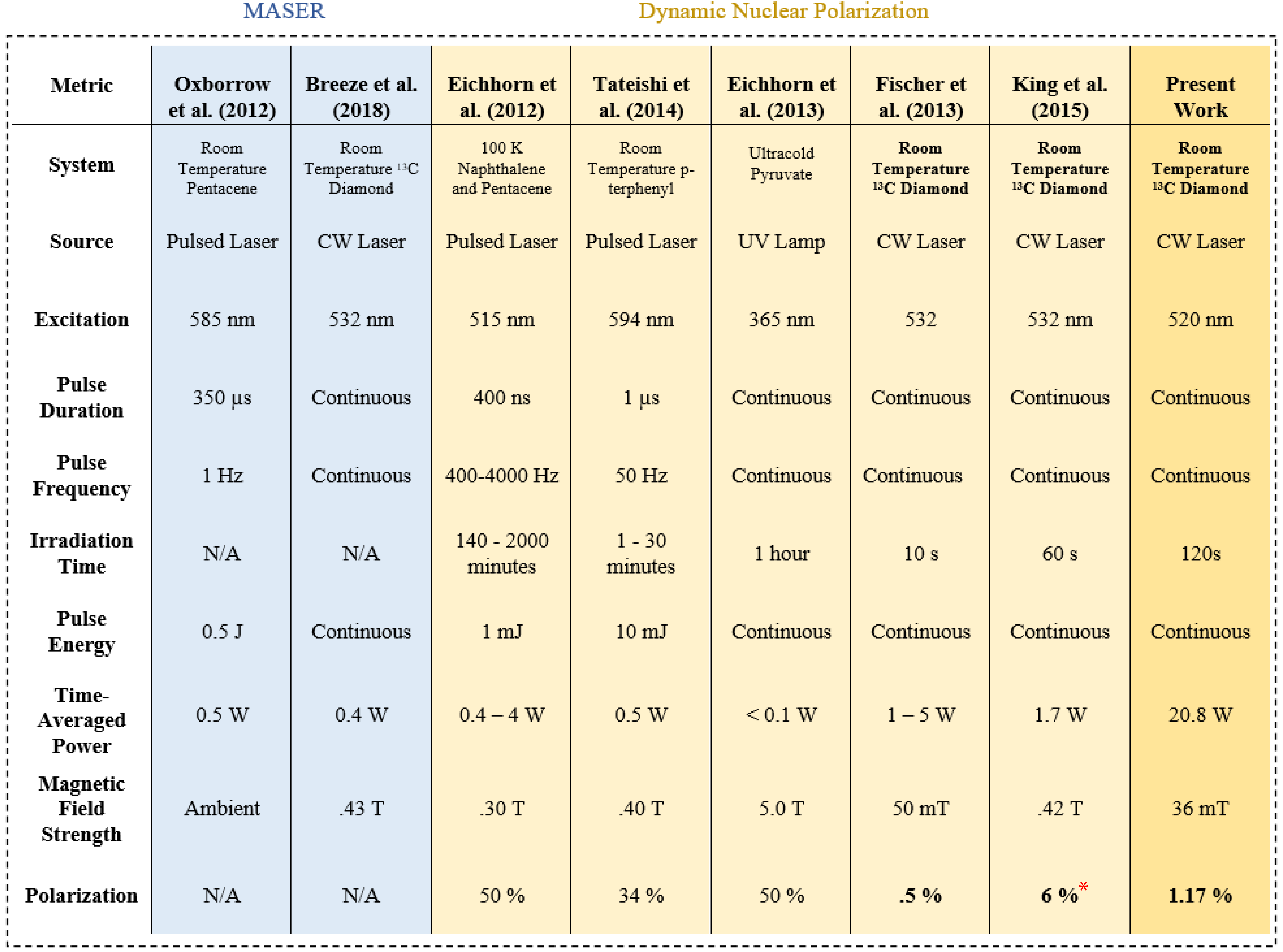}}
  \caption{Table comparing our optical system (\zfr{fig2}-\zfr{fig3}) to previous work in the literature for DNP and spin-based MASERs. Our multi-laser excitation and thermal management strategy is able to deliver much higher time averaged optical power compared to previous work. \red{${}^{\ast}$}We note that the polarization estimate in King et al.~\cite{King15} did not use an internal (thermal) reference.}
\zfl{figS6}
\end{figure*}

\section{Appendices}
\vspace{-4mm}
\begin{appendix}
\section{Enhanced optical excitation via "laser dome"}
\vspace{-4mm}
\zal{dome}
Here we detail the apparatus for high power optical illumination developed in this work. It consists of thirty lasers that deliver ${>}20$W of continuous, nearly isotropic, optical power to the sample. Previous experiments with NV centers, both for DNP and quantum sensing, were predominantly in the $\xn_e{<}$5W regime (see \zar{table}) and limited by sample heating~\cite{Wunderlich17b,Pagliero17,Ajoy20,Barry20}. There have been several reports of DNP with pulsed lasers~\cite{Tateishi14,Eichhorn13}. While possessing higher peak power, they entail long dead times (${\gg}T_{1e}$), where electrons are not being polarized. In contrast, the CW optical excitation here yields continuous $e$-repolarization, and is advantageous for DNP with electrons with broad ESR spectra~\cite{Ajoy18}. Furthermore, lower peak power injection here allows more efficient heat diffusion and time-averaged power before sample damage thresholds are reached.

The apparatus (shown in \zfr{fig2}) consists of a 3D printed carbon fiber dome-shaped structure  (``laser dome'') that houses the aforementioned 30 diode lasers ($\approx$0.8W each) delivered via multimode fibers (beam diameter ${\app}$4mm). More details of the construction with CAD models are presented in the supplementary material~\cite{SM}. We exploit relaxed requirements on optical mode quality,  polarization,  and stability necessary for $e$-polarization. This enables using an array of low-cost diode lasers to generate a high total optical power. The fibers are pressure fit into grooves that geometrically align towards the dome center (\zfr{fig2}(c)). The almost isotropic excitation pattern uniformly illuminates each of the sample facets and the exact beam arrangement is staggered for minimal overlap with the MW excitation coil (zoomed in \zfr{fig2}(d)). It also allows a significantly larger sample volume to be irradiated compared to previous approaches~\cite{Pagliero17,Ajoy20,Wolf15}, since an ${\app}4\pi$ solid-angle is illuminated. Ray-trace simulations depicted in \zfr{fig2}(e) help discern the best positions of laser fibers. Density of overlapping ray traces here serve as a proxy for the relative intensity of irradiation onto the sample (see supplementary material ~\cite{SM}).  The central portion of the sample sees excitation from multiple sources which, to an extent, compensates for attenuation through it~\cite{Acosta09,Mui08}. Although quantifying the exact penetration depth through the sample is experimentally challenging, we estimate it to be on the order of 0.1{-}0.15mm~\cite{Acosta09}. 

Large optical powers require the ability to mitigate sample heating.  We designed an \I{in-situ} heat exchanger that efficiently ejects heat while keeping the sample free from motion. \zfr{fig3}(a) describes its operation. The sample is held in a test tube surrounded by ${\app}4$mL water. Thermal energy injected into the diamond is rapidly dissipated to the water, which serves as a heat sink. The water is kept at a stable temperature by flowing cool nitrogen gas (-20$^{\degree}$C at inlet) across the test tube. The gas is delivered from slits built into the neck region of the laser dome (blue arrows in \zfr{fig3}(a)). Nitrogen flow rate is calibrated so the water temperature is ${\app}9^{\degree}$C when lasers are off (see supplementary material~\cite{SM} for more details).  Heat exchange exploits the excellent thermal conductivity of diamond (2200Wm$^{-1}$K$^{-1}$) and the large heat capacity of water for efficient thermal dissipation (red arrows in \zfr{fig3}(a)). The benefit of this relayed heat transfer strategy is that the cold gas does not contact the sample directly, and the sample volume can be enclosed -- an advantage for shuttling to high field for detection~\cite{Ajoyinstrument18}. 

Heat exchanger performance is found to be highly effective. \zfr{fig3}(b)-(c) depicts measured temperature buildup in the sample and surrounding fluid under 120s continuous irradiation at different optical powers. After 120s, the lasers are turned off, and the temperature dissipation is again monitored. Even at sustained 24W optical power (an intensity of ${\app}0.19$kW/cm$^2$) employing thirty lasers, the (asymptotic) steady-state temperature is less than $30^{\degree}$C and no sample damage is observed.  \zfr{fig3}(d) plots the steady-state temperatures for different powers. From the buildup rate, we estimate that ${\gtrsim}$50W power can be applied before system limits (related to water boiling) are reached. 

The strategy for high-power optical illumination and thermal management developed here is extensible to other systems, including organic triplet molecular systems, and UV generated non-persistent radicals~\cite{Eichhorn13,Capozzi17}. We envision applications to $e$-spin MASERs~\cite{Oxborrow12,Breeze18}, and quantum sensors with $e$-spin ensembles~\cite{Lesage12,Wolf15}, where reaching higher optical powers is the primary factor limiting magnetometer sensitivity~\cite{Barry20}.

\begin{figure}[t]
  \centering
  {\includegraphics[width=0.49\textwidth]{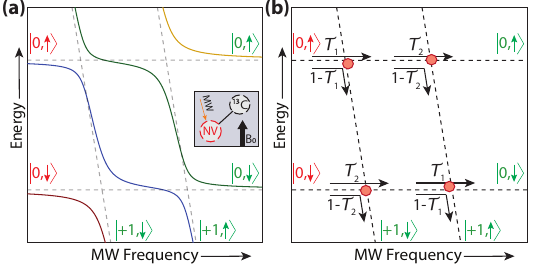}}
  \caption{\T{Single-spin ratchet. } (a) Landau-Zener level anti-crossings (LZ-LACs) for a system of an NV coupled to a single $\Cs$ nucleus (shown in inset) with $A^{\pll}{>}0$. Two pairs of energy gaps are visible, the large gaps being set by the electronic Rabi frequency. The two electronic manifolds are marked. Optical pumping resets the system polarization to the $m_s{=}0$ manifolds, and initially the system is unpolarized, yielding equal populations in either of the $\ket{0,\dw},\ket{0,\up}$ states. Hyperpolarization can be excited under a MW sweep from low-to-high frequency; optimal rate of the sweep $f_{\R{opt}}$ is determined by conditional traversals through the four LZ-LACs.  (b) LZ-LAC locations extracted from (a), and arranged in a checkerboard (\I{"Galton board"}~\cite{Pillai21}) of energy and frequency. Population bifurcation as the LZ-LACs are encountered under the MW sweep is captured by the tunneling probabilities $\mT_{1,2}$. Starting (red) and ending (green) nuclear states are marked. }
\zfl{figS1}
\end{figure}

\section{Comparison to previous work}
\vspace{-4mm}
\zal{table}
The table in \zfr{figS6} compares laser parameters of our work with other recent experimental studies focused on optically-pumped DNP~\cite{Eichhorn13, Tateishi14, King15, Fischer13} and $e$-spin MASERS~\cite{Oxborrow12}. Previous literature has explored both pulsed and CW light sources at a variety of different wavelengths. We see from this table that there are several advantages to our approach. First, we employ low-cost diode 520nm diode lasers that can be arrayed together to create a large continuous optical excitation power, as opposed to a singular, more expensive laser. This also exploits the very relaxed requirements on laser mode quality required for $e$-polarization. We are able to apply significantly higher time-averaged optical power than most previous experiments. Our data (\zfr{fig3}(b)-(d)) indicates that scaling up to even higher power can be relatively easily accomplished.  Pulsed laser systems can deliver larger peak power but are limited by slow repetition rates. They also make the sample more susceptible to damage from intense high power pulses~\cite{Oxborrow12}.  

\section{Derivation of polarization transfer rate for single-spin ratchet}
\vspace{-4mm}
\zal{ratchet}
We present here a derivation of the spin ratchet polarization rate (\zr{ratchet} in the main paper). For more details on the theory, assumptions and generalizations to large $N$, we refer the reader to Ref.~\cite{Elanchezhian21}. Consider \zfr{figS1}, where we show a single spin ratchet ($N{=}1$) with four Landau-Zener level anti-crossings (LZ-LACs) in the rotating frame of the MWs, here between states of the $m_s{=}0$ and  $m_s{=}1$ manifolds. Upon a MW sweep, there is a traversal of this cascaded LZ-LAC structure from left-to-right. We make the following assumptions that help simplify the theoretical evaluation of traversals, but which are reasonable under the regime of the experiments: \I{(i)} the LZ-LACs are assumed to be hit sequentially so that their effects can be evaluated individually, \I{(ii)} the nuclear populations are assumed to start in the $m_s{=}0$ manifold and bifurcate "down" or "right" upon encountering the LZ-LACs, and \I{(iii)} NV electronic (re)polarization is assumed to happen far away from the exact LZ-LAC points. 

Consider a single traversal from left-right through the $N{=}1$ LACs in \zfr{figS1}, starting from populations restricted to the $m_s{=}0$ manifold and ending with NV repolarization. The probabilities of the possible population evolutions are then,
\bea
\mP(\dw\: \rightarrow\: \dw)&=&\left(1-\mT_{2}\right)+\mT_{2} \mT_{1} \non\\
\mP(\dw\: \rightarrow\: \up)&=&\mT_{2}\left(1-\mT_{1}\right) \non\\
\mP(\up\: \rightarrow\: \dw)&=&\mT_{2}\left(1-\mT_{1}\right)+2 \mT_{1}\left(1-\mT_{1}\right)\left(1-\mT_{2}\right) \non\\
\mP(\up\:\rightarrow\: \up)&=&\mT_{1}\mT_{2}+\mT_{1}^{2}\left(1-\mT_{2}\right)+\left(1-\mT_{1}\right)^{2}\left(1-\mT_{2}\right) , \non
\eea
where $\mT_{1,2}$ here refer to the tunneling probabilities (\zfr{figS1}(b)), and depend on the adiabaticity of the traversal through the respective LZ-LAC; for example, $\mT_{1,2}=\exp(\vxe_{1,2}^2/f_r\mB)$. Indeed, each term in the expressions above can be thought of as referring to a different trajectory through the LZ-LAC structure (schematically shown in \zfr{figS1}(b)).  For instance, there are two paths that constitute the term, $\mP(\dw\: \rightarrow\: \dw)$, corresponding to the probabilities of $\ket{0,\dw}{\rt}\ket{0,\dw}$ and $\ket{0,\dw}{\rt}\ket{+1,\dw}$ respectively. Now, in order to determine the nuclear hyperpolarization, we evaluate the difference in populations between the nuclear states at the end of the sweep, 
\bea
P&=&\lsb \mP(\dw\: \rightarrow\: \dw) + \mP(\up\: \rightarrow\: \dw)\rsb -\lsb \mP(\dw\:\rightarrow\: \up) + \mP(\up\:\rightarrow\: \up)\rsb\non\\
&=&\left(1-\mT_{2}\right)\left[1-\left(2 \mT_{1}-1\right)^{2}\right]
\zl{eq3} .
\eea
Ultimately, \zr{eq3} illustrates that hyperpolarization develops as a result of the differential adiabaticity of the traversals through the pairs of LZ-LACs conditioned on the nuclear state. The net hyperpolarization developed in time $T$ has the form,
\beq
P=\left[1-\exp \left(\frac{-c_e{\eta}_{e}}{f_{r}}\right)\right] \cdot T f_{r} \cdot\lsb\left(1-\mT_{2}\right)\lcb 1-\left(2 \mT_{1}-1\right)^{2}\rcb\rsb,
\zl{eq4}
\eeq
where the first term encapsulates the starting electron polarization, the second term gives the total sweeps in time $T$, and the last term is \zr{eq3}. We therefore refer to the mechanism as being a "spin-ratchet", since each MW sweep event can be thought of as performing work to transfer a finite amount of polarization from the electronic spin to the directly coupled nuclei. While \zr{ratchet} is derived from the $N{=}1$ case displayed in \zfr{figS1}, it acts as a good approximation even at large $N$ such as in the system in \zfr{fig1}(a). This generalization to large $N$ is further detailed in Ref. ~\cite{Elanchezhian21}. Here, one can show that the system comprises a cascaded structure of $2^{2N}$ LZ-LACs, and evolution through them can be theoretically evaluated by mapping the dynamics to the operation of a "Galton board"~\cite{Elanchezhian21}.
\end{appendix}
\bibliography{main.bbl}
\pagebreak

\clearpage
\onecolumngrid
\begin{center}
\textbf{\large{\textit{Supplementary Information} \\ \smallskip
Rapidly enhanced spin polarization injection in an optically pumped spin ratchet}}\\
\hfill \break
\smallskip
A. Sarkar$^{1},^\ast$, B. Blankenship$^{1},^\ast$, E. Druga$^{1}$, A. Pillai$^{1}$, R. Nirodi$^{1}$, 
S. Singh,$^{1}$ A. Oddo$^{1}$,  P. Reshetikhin,$^{1}$ A. Ajoy$^{1,2}$\\
${}^{1}$\I{{\small Department of Chemistry, University of California, Berkeley, Berkeley, CA 94720, USA.}}\\
${}^{2}$\I{{\small Chemical Sciences Division Lawrence Berkeley National Laboratory,  Berkeley, CA 94720, USA.}}\\

\end{center}

\twocolumngrid

\beginsupplement
\setcounter{section}{0}
\setcounter{page}{1}
\section{Hyperpolarization setup}
\vspace{-4mm}
\zfr{figS2} shows the overall hyperpolarization setup, containing the lasers employed for DNP, along with thermal management systems and a Helmholtz coil for application of the polarization field. The setup consists of a three-sided panel design, each with total edge length 18in.  and constructed in a compact manner for positioning under a 7T magnet. The laser diodes were mounted on the interior panels, and driver modules on the exterior panels as shown in \zfr{figS2}(a)-(b). This allows the use of short 0.5m fibers while preventing tight bends. 

The panels were designed in a manner that allows the laser diodes and drivers to be easily accessible. Each exterior panel holds up to 12 driver modules (Lasertack PD-01289), and each interior panel holds the corresponding Lasertack 520nm laser diodes. The structure can mount laser diodes and modules on up to four sides (allowing a total of 32 lasers). These panels are fastened onto 8020 aluminum frames and can easily screw on or off without having to remove any other parts of the structure. The laser diodes are electrically connected to the driver modules by snap-on connectors. For our experiments we mounted a total of 30 lasers ($\app$0.8W each) onto three panels. 

An Arduino Leonardo microcontroller mounted on each panel, controls the laser diodes. Each microcontroller connects to a central computer via USB serial communication. This permits the ability to turn on or off any combination of lasers with individually set times and duty cycles that can be dynamically reprogrammed.

A large 12in. Helmholtz coil surrounds the dome at the center of the setup. The dome and coil require precise vertical positioning relative to the 7T magnet that is difficult to calibrate \I{ex-situ}. To ease in-situ calibration we constructed a multilevel mount connected to a plexiglass plate cover by bolts that can be adjusted to change the vertical positions of the coil and dome simultaneously. This structure is designed such that all of the components could be fabricated in-house using an entry level laser-cutter or 3D printer.

We took a multi-tiered approach to thermal management. There is a need to cool both the sample and the laser diodes themselves to ensure optimal performance. We focus our discussion here to the laser diode cooling, while details of the sample cooling (heat exchanger) are presented in Sec. \zsr{heatex}. A Peltier cooler (TE Inc. TE-63-1.0-1.3) is attached to the base of each diode to maintain a temperature at or below 25$^{\circ}$C inside the laser (\zfr{figS2}(d)). This maintains excitation wavelength and power at 520nm and $\app$0.8W respectively, even under continuous operation. To prevent overheating at the laser panel interface, a cooling block fed with -20$^{\circ}$C nitrogen gas is attached to the base of the device. The slightly heated gas at the exit port of the cooling block is then sent to a copper tube running through each of the laser heads for additional cooling.

We subjected the hyperpolarization device to a series of stress tests to ensure that it could meet the desired operating conditions. One such test involved running the lasers at full power ($\app$0.8W optical output each) for one minute on and three minutes off for 6 hours, by which point we assumed that the system would have reached thermal equilibrium with its surroundings. While conducting this test we closely monitored temperature across the device via infrared cameras. The laser temperature was found to be under 30$^{\circ}$C even during sustained operation.

\begin{figure}[t]
  \centering
  {\includegraphics[width=0.5\textwidth]{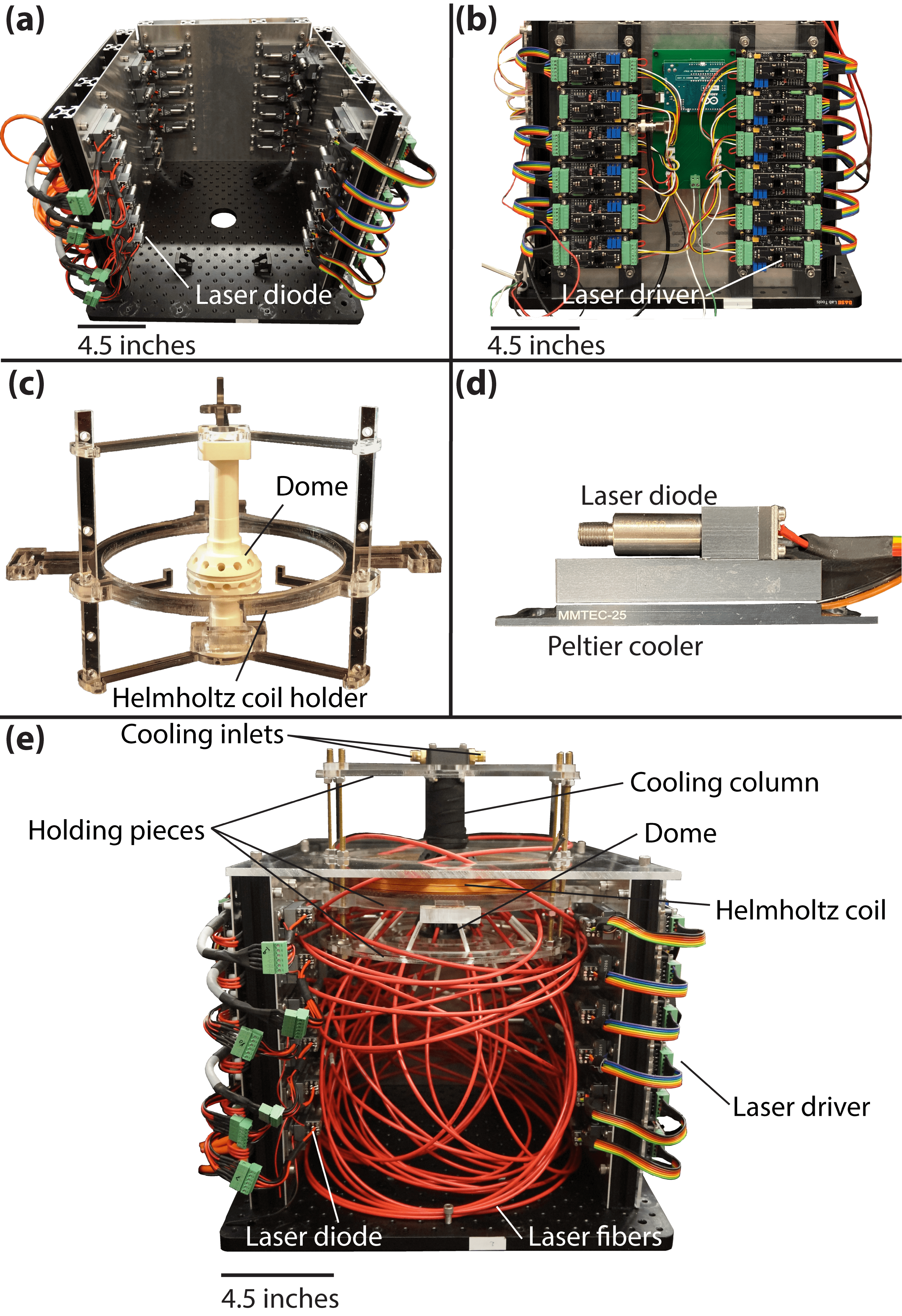}}
  \caption{\T{Construction of the laser excitation setup. } (a) Photograph of the laser excitation setup consisting of 30 individual lasers placed on three faces (walls) of a cubical (12in) box, built out of 1/4in sheet aluminum. Fibers here are shown removed for clarity. The laser heads (see (d)) occupy the inner faces, while the drivers are placed on the outer faces. (b) Picture of an outer face showing 12 mounted laser drivers. They are interfaced to power by means of a custom centrally located PCB. (c) Photograph of supporting structure laser-cut out of plexiglass which supports the laser dome and the Helmholtz coil. It also ensures central alignment of the entire structure and permits rapid insertion and extraction of the sample from it. (d) Single laser diode (Lasertack). It consists of a fiber coupling attachment where a multimode fiber is screwed on. Its lower surface consists of a Peltier cooler for temperature regulation. We employ 30 such lasers, each with an output power (after the fiber) of $\app$0.8W. (e) Photograph of the fully assembled apparatus with fibers connected, and the laser dome mounted. The fibers (each 0.5m long) are coiled in a symmetric fashion to ensure tight laser packing. All structural support is provided by the plexiglass structure in (c). The 3D printed sample cooling column at the neck of the dome (see \zfr{fig2} of the main paper) is visible in the top portion of the picture along with the plexiglass supports.}
\zfl{figS2}
\end{figure}

\begin{figure}[t]
  \centering
  {\includegraphics[width=0.49\textwidth]{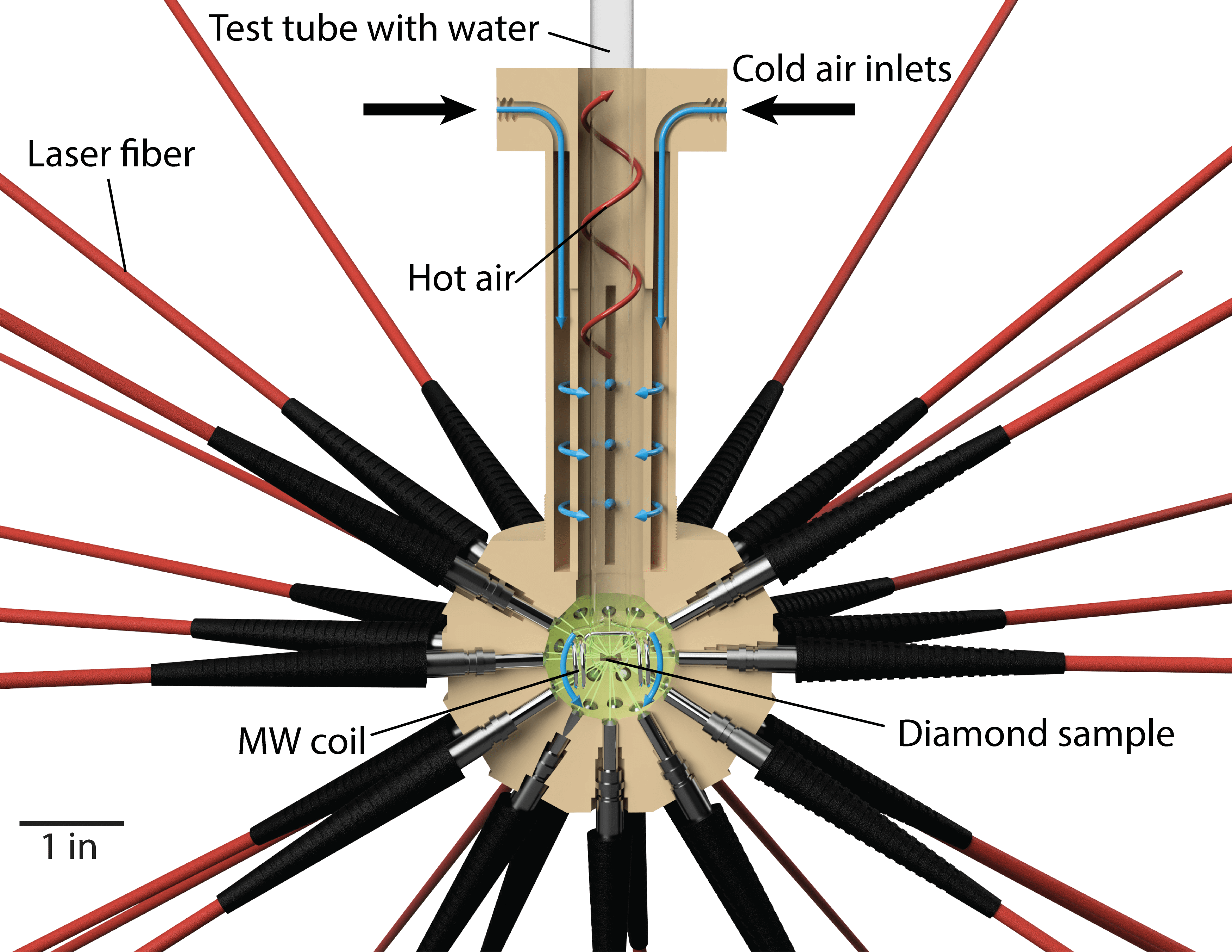}}
  \caption{\T{Laser dome apparatus for multilaser excitation. } Schematic figure showing the cross-sectional view of the laser \I{“dome”} apparatus employed for high-power optical excitation (\zfr{fig2}(b) of the main paper). The device houses 30 impinging laser beams while simultaneously being able to extract heat from the sample. Body (yellow) of the ``dome’’ is a 3D printed structure with slots (grooves) from which the laser excitation is delivered (see \zfr{figS4}). Lasers here are diode lasers and excitation is carried by multi-mode optical fibers that are pressure fit into the grooves.  Optical fibers are arranged in an approximately spherical arrangement in three layers. At the center of the dome is the MW (split) coil that is used to excite nuclear hyperpolarization. Sample is carried in a 8mm test-tube inserted through a cavity at the top of the {``neck’’} of the dome, and running through the length of the device. Sample is immersed in water, and the neck region serves as the heat exchanger to extract thermal energy from the sample (see \zfr{fig3} of the main paper). Top of the 3D printed structure here has lateral slots through which cold nitrogen gas is inserted (blue arrows) into the central cavity. This air cushion cools the walls of the test tube and through it the water surrounding the sample; hot air exits from the top of the dome (red arrows).}
\zfl{figS3}
\end{figure}

\begin{figure}[t]
  \centering
  {\includegraphics[width=0.49\textwidth]{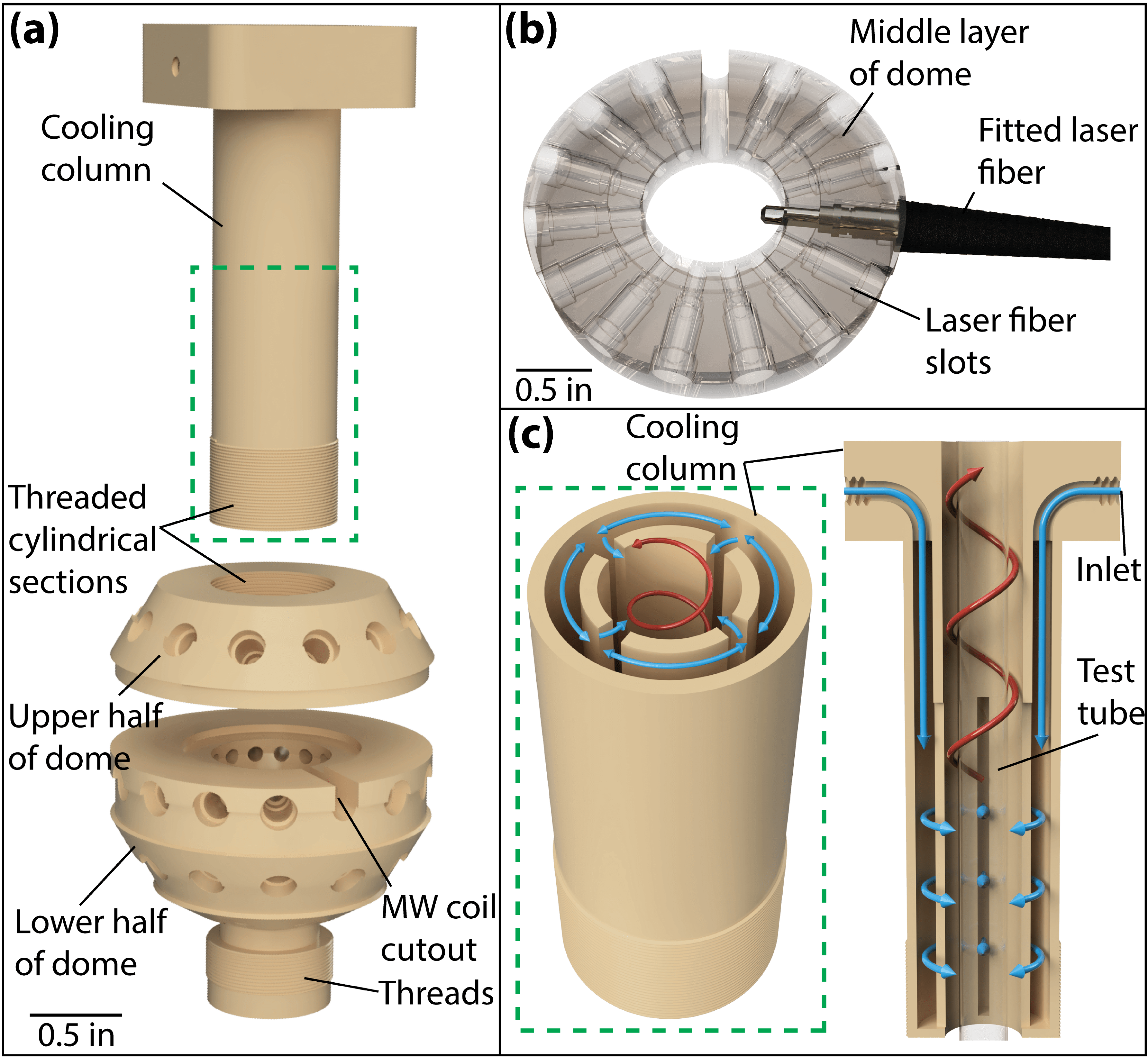}}
  \caption{\T{Individual parts of the laser dome apparatus} in \zfr{figS2}. (a) CAD rendering of 3D printed structure showing constituent parts that are screwed together (via 3D printed threads). These are marked individually as: the neck region that serves as the heat exchanger, and the top and bottom halves of the dome which host a transverse cavity through which the coaxial cable for MW delivery is inserted.  The bottom half houses threads that screw into a Thorlabs CP02 mount held on the laser cut plexiglass structure (see \zfr{figS2}(c)) that carries the Helmholtz coils, ensuring central alignment. (b) CAD rendering of a single layer of the central dome showing the pressure-fit attachment of the laser fiber into it. Tight attachment to the 3D printed dome is arranged for here by friction, with no screws or supporting structures used. This greatly enhances laser packing density. (c) Detailed view of heat exchanger operation. Cross sectional (top and side) views of the cooling column at the neck of the laser dome are shown to illustrate heat extraction.  Evident is the central 9mm cavity from where the glass tube carrying the sample is shuttled. 3D printed transverse slots permit the ability to guide the flow of nitrogen gas onto and from the sample tube. Blue and red arrows here denote the flow of cool and warm gas respectively. Flowing nitrogen gas cools the water surrounding the diamond sample and actively extracts heat from it (see \zfr{fig3} of the main paper).}
\zfl{figS4}
\end{figure}

\section{"Laser Dome": design and construction}
\vspace{-4mm}
The design of the laser dome (\zfr{fig2} of the main paper) sought to achieve two complementary objectives: \I{(i)} mount the maximum number of laser fibers and hence deliver the maximum power to the sample, and \I{(ii)} arrange the lasers so that the beams excite the NV centers approximately isotropically. To accomplish this, we designed a polyhedral structure of four stacked, coaxial rings, with the top portion of the dome occupied by a cooling column (see \zfr{figS3}). The sample is placed in a test tube in the central cavity of the dome and the fibers are mounted into the dome 4mm from the sample center. 

The absence of a collimator greatly reduces the real estate required for each laser, and ensures optimal divergence to illuminate the sample. We are therefore able to fit 8-14 lasers on each 3D printed ring. \zfr{figS4}(b) shows one such ring with an added slot for the microwave coil. The 3D printed cavities for fibers are designed in a manner that the unit normal vector of all the fibers points towards the dome centroid where the sample is positioned. Each adjacent fiber tip has $\lesssim$1.5mm separation from its neighbor. The dome was 3D printed in carbon fiber composite material, which was chosen for its high melting point. During fabrication, the layer thickness was varied to achieve the required smoothness of the structure, ensuring tight fits for the fibers. 

The MW coil is tightly fit into the laser dome without touching the test tube or inner walls of the dome. It is designed in a manner that minimizes obstruction with the laser beam paths.

\section{Sample Thermal Management}
\vspace{-4mm}
\zsl{heatex}
Here we describe more technical details of the sample cooling heat exchanger design. Large optical powers make handling sample heating a necessity and an important technical challenge. The 3D printed laser dome encapsulates a cooling column at its top portion to enable efficient sample cooling. This cooling system exploits three key features: \I{(i)} excellent thermal conductivity of diamond that transmits the injected heat to the surrounding water, \I{(ii)} high heat capacity of the water, and \I{(iii)} cold air flow that works to eject heat from the water. Overall, this creates an efficient and hyperpolarization compatible heatsink around the diamond. Importantly, the sample test tube itself has no liquids flowing into it, and hence it can be mechanically shuttled for NMR measurements~\cite{Ajoyinstrument18}.

The sample cooling column (\zfr{figS3}-\zfr{figS4}) consists of two concentric slotted cylindrical shells. Cooled nitrogen (at -20$^{\circ}$C) is fed into ports from the top of the cylindrical column at 48 cfm to chill the water surrounding the diamond. Similar to the operation of a Dyson bladeless fan, the slots in the cylindrical shells create vortex currents which increase the characteristic length of heat transfer. Additionally, this design increases turbulence within the chamber, which better dissipates heat within the gaseous medium. The top of the column has an orifice for warmer gases to escape; this is also from where the sample test tube is shuttled for $\Cs$ NMR~\cite{Ajoyinstrument18}.

Our experimental observations of the heat exchanger performance are presented in \zfr{fig3} of the main paper. In experiments, a K-type thermocouple is used to record system temperature under 120s  of continuous illumination with an increasing number of lasers. We visually observe small yet complicated microbubbling at the boundary of the diamond. Fortunately , the bubbling has a neglible effect on heat exchange. Overall, we are able to keep the system steady state temperature under 30$^{\degree}$C even under the sustained high-power illumination.

\section{Optical Simulation}
\vspace{-4mm}
Our experiments are carried out on a 4x4x1mm diamond sample. Using the exact position of the fibers, we employed COMSOL to carry out a simple optical simulation based on laser ray tracing from the laser fiber tip positions in the dome, passing through air (refractive index $n{=}1$), glass ($n{=}1.6$), water ($n{=}1.33$) and ultimately the diamond ($n{=}2.54$) (\zfr{fig2}(e) of the main paper). We included effects of refraction, and the intrinsic 6° laser beam divergence. Attenuation was not included; attenuation in air, glass (absorption coefficient $k{=} 0.34$ m$^{-1}$ ), and water ($k{=} 0.045$ m$^{-1}$) is negligible. The optical simulation (\zfr{fig2}(e) of the main paper) shows an overlapping pattern of ray traces indicating a concentration of light at the center of the diamond. Diamond has a high absorption coefficient of 12.7 mm$^{-1}$ and penetration depth of 0.0787 mm; this regime overcomes some of the potential attenuation losses due to the overlapping rays produced by the lasers.

\begin{figure}
  \centering
  {\includegraphics[width=0.49\textwidth]{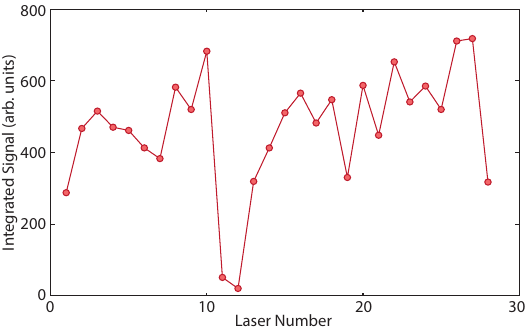}}
  \caption{\T{Characterization of the hyperpolarizaton signal} from each laser. Panel shows the signal obtained from each laser operated individually.  Lasers are numbered from top-to-bottom and clockwise. We find an approximately homogeneous generation of hyperpolarization with slight variations stemming from geometric effects in the manner in which the laser beam strikes the sample. Two laser diodes (number 11 and 12) show worse performance because they have degraded with time and have a lower power output. }
\zfl{figS5}
\end{figure}

\section{Characterization of laser homogeneity}
\vspace{-4mm}
In this section, we describe experiments to characterize and optimize the homogeneity of the laser excitation on the diamond sample. Our goals are to maximize the uniformity of the optical illumination in order to ensure that every part of the sample sees approximately the same laser excitation intensity (here $\app$0.8W per laser). These measurements will be employed in order to normalize the effective laser intensities in the experiments performed in \zfr{fig4} of the main paper. Optical inhomogeneity itself can arise from factors including:

\I{i.}	Misplacement of the laser. From the vantage point of a few lasers excitation slots, the MW coil casts a shadow on the sample. Lasers mounted on these positions therefore strike the sample with their intensity greatly suppressed. The exact positions depends on the exact MW coil employed in the experiment. We find that for a typical split coil design, 30 lasers can easily be accommodated with minimal loss. 

\I{ii.}	Off-center MW coil. The 3D printed slots in the laser dome ensure that fibers placed into them are naturally aligned to the exact geometric center of the dome structure. This assumes, however, that the MW coil itself is centrally located in the dome cavity in order to minimize overlap with the laser beams. 

\I{iii.}	Total internal reflection. Depending on the size, shape and orientation of the sample, some of the lasers are more effective because optical beams from them undergo total internal reflection within the diamond.

Hyperpolarization levels obtained from each laser applied individually to the sample provide the best means to characterize the homogeneity of the optical excitation. We carry out such DNP experiments with each laser illuminated for 20s, and measure the $\Cs$ hyperpolarized NMR signal under pulsed spin lock for 20s. The resulting integrated signals are then correlated with position on the dome. Any inhomogeneities due to coil or laser misplacement can then be easily identified. It is evident from \zfr{fig2}(e) of the main paper that hyperpolarization can be arranged to be homogenous to a good degree even for the 28 applied lasers. Two lasers in \zfr{figS5} had degraded in performance and were at reduced intensity. The experiments also reveal that the maximum hyperpolarization intensity arises, somewhat counterintuitively, for lasers striking the sample at the diagonal top positions.

\end{document}

%% file: Commands3.tex
\newcommand{\vxe}{\varepsilon}
\newcommand{\tm}{{\text -}}

\newcommand{\xg}{\gamma}

\newcommand{\xn}{\eta}
\newcommand{\xk}{\kappa}

\newcommand{\xo}{\omega}

\newcommand{\pp}{\perp}
\newcommand{\app}{\approx}
\newcommand{\dw}{\downarrow}
\newcommand{\up}{\uparrow}

\newcommand{\Cs}{{}^{13}\R{C}}

\newcommand{\degree}{^{\circ}}

\newcommand{\mB}[0]{\mathcal{B}}

\newcommand{\xD}{\Delta}

\newcommand{\xO}{\Omega}

\newcommand{\mP}[0]{\mathcal{P}}

\newcommand{\fr}[2]{\frac{#1}{#2}}

\newcommand{\mH}[0]{\mathcal{H}}

\newcommand{\mT}[0]{\mathcal{T}}

\newcommand{\rt}{\rightarrow}

\newcommand{\beq}{\begin{equation}}
\newcommand{\eeq}{\end{equation}}
                  
\newcommand{\benum}{\begin{enumerate}}
\newcommand{\eenum}{\end{enumerate}}
                    
\newcommand{\bit}{\begin{itemize}}
\newcommand{\eit}{\end{itemize}}

\newcommand{\bea}{\begin{eqnarray}}
\newcommand{\eea}{\end{eqnarray}}

\newcommand{\bev}{\begin{verbatim}}
\newcommand{\eev}{\end{verbatim}}

\newcommand{\non}{\nonumber}

\newcommand{\lsb}{\left[}
\newcommand{\rsb}{\right]}
\newcommand{\lcb}{\left\{}
\newcommand{\rcb}{\right\}}

\newcommand{\pll}{\parallel}

\newcommand{\T}[1]{\textbf{#1}}
\newcommand{\I}[1]{\textit{#1}}
\newcommand{\R}[1]{\textrm{#1}}

\newcommand{\zl}[1]{\label{eqn:#1}}
\newcommand{\zr}[1]{Eq.\,(\ref{eqn:#1})}
\newcommand{\zfl}[1]{\protect\label{fig:#1}}
\newcommand{\zfr}[1]{\figurename\,\ref{fig:#1}}

\newcommand{\zsl}[1]{\label{sec:#1}}
\newcommand{\zsr}[1]{\!\ref{sec:#1}}
\newcommand{\zal}[1]{\label{app:#1}}
\newcommand{\zar}[1]{Appendix \ref{app:#1}}

\newcommand{\ket}[1]{\left\vert{#1}\right\rangle}

\newcommand{\ba}{\left\{ \begin{array}{lr}}
\newcommand{\ea}{\end{array}\right.}

\newcommand{\blist}[1]{
 \begin{list}{#1}
 \begin{align}
	 arrow
 \end{align}
 $\checkmark\star
  { \setlength{\itemsep}{3pt}
     \setlength{\parsep}{2pt}
     \setlength{\topsep}{3pt}
     \setlength{\partopsep}{0pt}
     \setlength{\leftmargin}{1em}
     \setlength{\labelwidth}{1em}
     \setlength{\labelsep}{0.5em} } }
\newcommand{\elist}{
  \end{list}  }

\DeclareMathSymbol{\vartheta}{\mathalpha}{letters}{"12}
\DeclareMathSymbol{\theta}{\mathalpha}{letters}{"23}
\DeclareMathSymbol{\phi}{\mathalpha}{letters}{"27}
\DeclareMathSymbol{\varphi}{\mathalpha}{letters}{"1E}

\newcommand{\bef}
{
\begin{figure}[htbp]
\centering
}

\newcommand{\eef}{\end{figure}}
